\documentclass[preprint,12pt]{elsarticle}

\usepackage{amsmath}    
\usepackage{graphicx}   
\usepackage{verbatim}   
\usepackage{color}      
\usepackage{subfigure}  
\usepackage{hyperref}   

\journal{Physics Letters B}

\begin{document}

\sloppy

\begin{frontmatter}

\title{Single flavor staggered fermions} 
\author{Christian Hoelbling}
\ead{hch@physik.uni-wuppertal.de}
\address{Bergische
  Universit\"at Wuppertal, Gaussstr.\,20, D-42119 Wuppertal, Germany}
\date{\today}

\begin{abstract}
Based on recent work by Adams, I construct a lattice fermion operator
that fully lifts the staggered flavor degeneracy. The resulting
operator is of Wilson type but smaller by a factor of 4, better
conditioned and contains 3 instead of 15 doublers. It is further
suggested that this operator may be used as a candidate kernel
operator to an overlap construction. Prospects for practical
applications and potential problems of the new discretizations are
briefly discussed.
\end{abstract}

\end{frontmatter}

When regularizing a massless continuum Dirac operator on the lattice,
one is faced with the fermion doubling problem
\cite{Karsten:1980wd,Nielsen:1980rz}. In addition to the single,
physical fermion flavor, 15 doubler fermions appear at the edges of
the Brillouin zone. Traditionally there have been two mutually
exclusive strategies to ameliorate or solve this problem: One can
either remove an exact fourfold degeneracy of the naive discretization
\cite{Kogut:1974ag,Banks:1975gq,Susskind:1976jm} and reduce the
doubling problem to 3 doubler species, or, alternatively, one can
introduce a momentum dependent mass term to lift the degeneracy of the
15 doubler species with the physical one \cite{Wilson:1975xx}.

In this paper I give an explicit construction of a single flavor
fermion operator that combines the above mentioned two approaches. It
is based on recent work by Adams \cite{Adams:2010gx} where the two
flavor case has been discussed. The construction involves adding a
momentum dependent mass term as in \cite{Wilson:1975xx}, but it starts
from the staggered operator \cite{Susskind:1976jm} rather than the
naive fermion operator. The resulting operator will be Wilson-like (in
particular it will break chiral symmetry and require an additive mass
renormalization) but smaller by a factor of 4 and with a better
condition number and only 3 doubler fermions. It is further suggested,
that this Wilson-like operator might be a suitable overlap kernel
operator.

The starting point of our construction is the massless staggered
fermion operator\cite{Susskind:1976jm}
\begin{equation}
D_\text{st}=\eta_\mu D_\mu, \qquad D_\mu=\frac{1}{2}\left(V_\mu-V_\mu^\dag\right)
\end{equation}
with $\left(\eta_\mu\right)_{xy}=(-1)^{\sum_{\nu<\mu}x_\nu}\delta_{x,y}$ and
$\left(V_\mu\right)_{xy}=U_\mu(x)\delta_{x+\hat{\mu},y}$. This
operator obeys a remnant chiral symmetry
\begin{equation}
\left\{D_\text{st},\epsilon\right\}=0
\end{equation}
where $\epsilon_{xy}=(-1)^{\sum_\mu x_\mu}\delta_{x,y}$. In the
spin-flavor interpretation of staggered fermions
\cite{Gliozzi:1982ib,KlubergStern:1983dg}, $\epsilon$ is identified with
$\epsilon=(\gamma_5\otimes\xi_5)$.

Following \cite{Adams:2010gx}, we first introduce a Wilson-like flavor
dependent mass term
\begin{equation}
M_1=\epsilon\eta_5C
\label{m1}
\end{equation}
with
$\left(\eta_5\right)_{xy}=\left(\eta_1\eta_2\eta_3\eta_4\right)_{xy}=(-1)^{x_1+x_3}\delta_{x,y}$
and a Laplacean-like term $C=(C_1C_2C_3C_4)_\text{sym}$ that is the
fully symmetrized product of the
$C_\mu=\frac{1}{2}\left(V_\mu+V_\mu^\dag\right)$. It is important to
note that the spin-flavor structure of $M_1$ is
\begin{equation}
M_1\sim(\mathbf{1}\otimes\xi_5)+\mathcal{O}(a)
\label{sfm1}
\end{equation}
Additionally, $M_1$ has two crucial properties: it is hermitian and
commutes with $\epsilon$. Both of these follow straightforwardly from the
definitions.

\begin{figure}
\subfigure[Operator (\ref{adop}) proposed in \cite{Adams:2010gx}]{
\label{figopsad}
\includegraphics[width=0.3\textwidth]{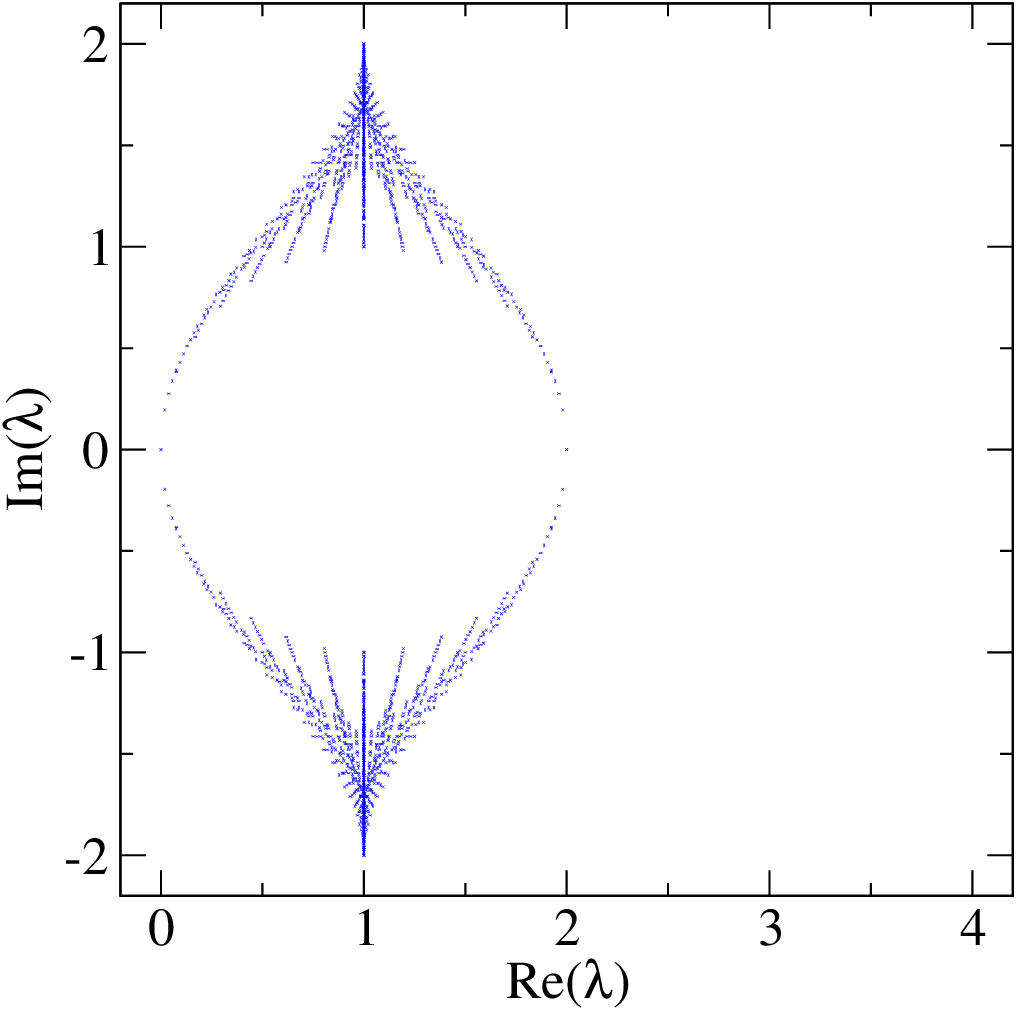}}
\subfigure[Unsymmetrized operator (\ref{myop})]{
\label{figopsmy}
\includegraphics[width=0.3\textwidth]{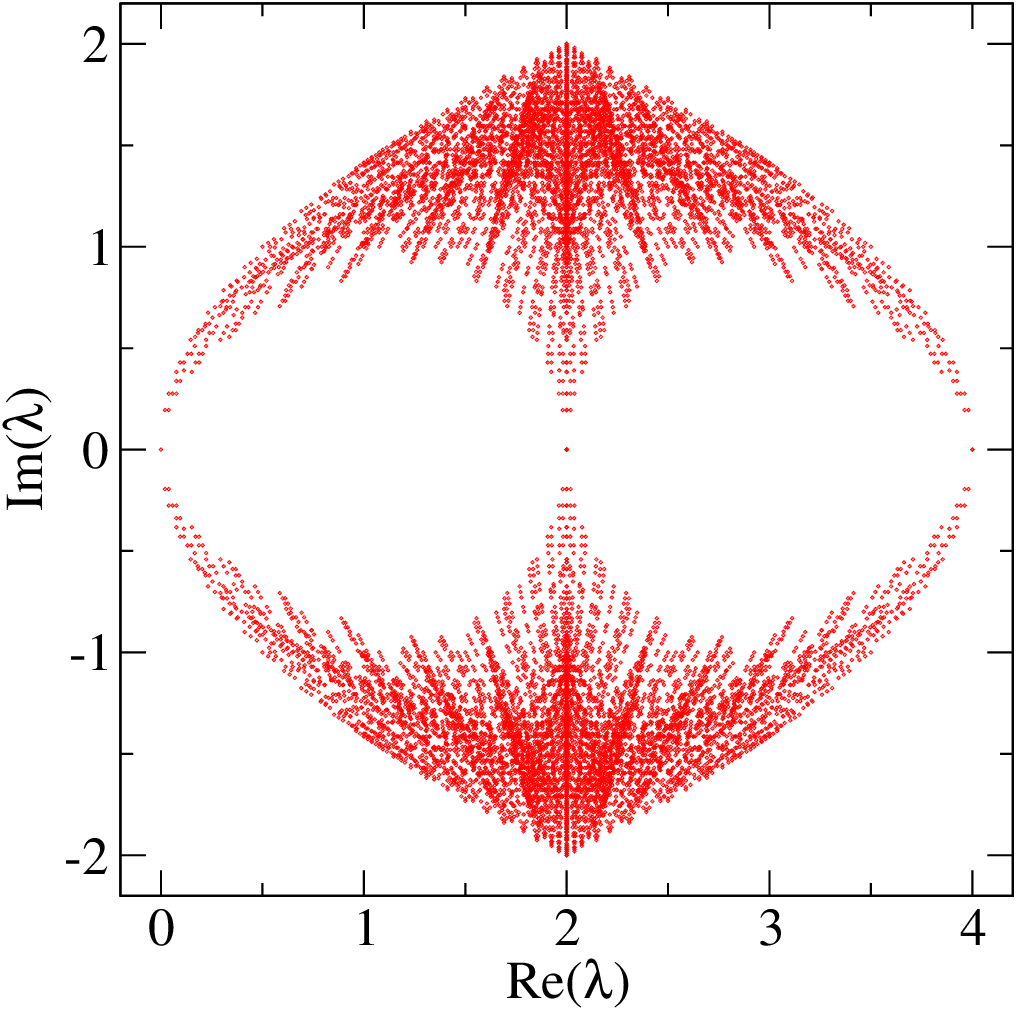}}
\subfigure[Symmetrized operator (\ref{myops})]{
\label{figopsmys}
\includegraphics[width=0.3\textwidth]{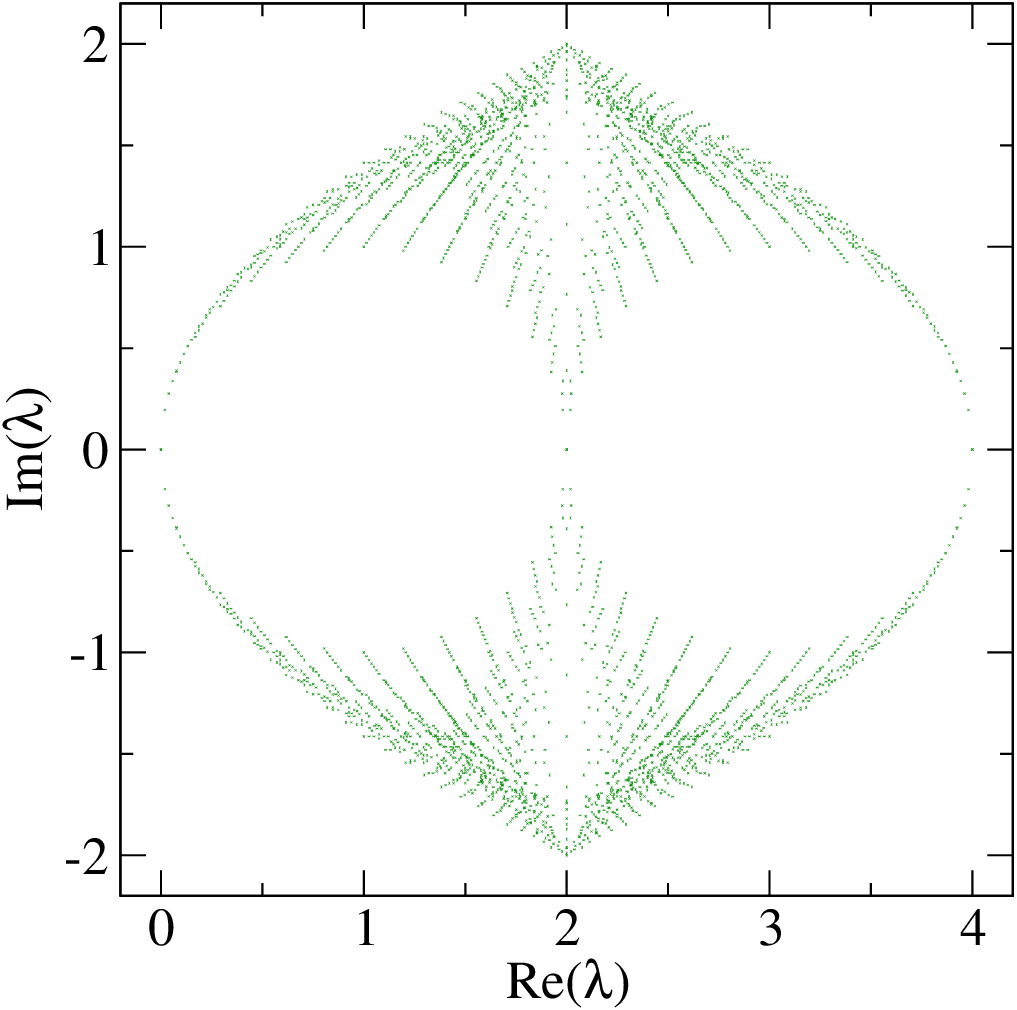}}
\caption{
\label{figops}
Spectrum of various Wilson-type operators with $r=1$ and $m_0=0$ in the
free field case on a $32^4$ lattice.}
\end{figure}

Using these properties, we can immediately see that the modified staggered operator
\footnote{Note that we have also added a mass term $r+m_0$ in order
  to shift the physical part of the spectrum to the correct position.}
\begin{equation}
D_\text{A}(m_0)=D_\text{st}+r\left(1+M_1\right)+m_0
\label{adop}
\end{equation}
with the Wilson-like parameter $r$ fulfills a $\gamma_5$-hermiticity
like condition $D_\text{A}(m_0)\epsilon=\epsilon
D^\dag_\text{A}(m_0)$. Consequently, its non-real eigenvalues appear in
complex conjugate pairs which ensures positivity of the determinant
for a suitable choice of $m_0$. Due to its spin-flavor structure
(\ref{sfm1}), the addition of $M_1$ in (\ref{adop}) will spread out
the spectrum in the real direction, giving modes a mass term according
to their approximate flavor chirality (cf. fig.~\ref{figopsad}). It
was demonstrated in \cite{Adams:2010gx} that this operator is a
suitable overlap kernel. The resulting overlap operator obeys an index
theorem with two fermion flavors \cite{Adams:2010gx,Adams:2009eb}.

The fact that one is left with two fermion flavors originates in the
dimension two of the positive and negative flavor chirality subspaces
in four space-time dimensions. In order to lift this remaining
degeneracy, an additional operator is needed, which differentiates
between flavors of the same chirality. In the flavor Clifford algebra,
the natural candidates are the matrices
$\sigma_{\mu\nu}=i\xi_\nu\xi_\mu$. The $\sigma_{\mu\nu}$ commute
with $\xi_5$ and can therefore simultaneously be
diagonalized. Furthermore, $\sigma_{\mu\nu}$ has one eigenvalue $1$
and one $-1$ in both the positive and negative chirality subspace.
Therefore, one can choose a common diagonal basis where
\begin{equation}
\xi_5=\text{diag}(1,1,-1,-1)
\qquad
\sigma_{\mu\nu}=\text{diag}(1,-1,-1,1)
\label{fs}
\end{equation}
In order to fully lift the flavor degeneracy, we therefore would need
an additional ``mass term'' with the flavor structure
$(\mathbf{1}\otimes\sigma_{\mu\nu})+\mathcal{O}(a)$.

A candidate for such an additional term that has the correct flavor
structure is given by
\begin{equation}
M_2^{\mu\nu}=i\eta_{\mu\nu}C_{\mu\nu}
\label{m2}
\end{equation}
with
\begin{equation}
\begin{split}
\left(\eta_{\mu\nu}\right)_{xy}&=-\left(\eta_{\nu\mu}\right)_{xy}=(-1)^{\sum_{i=\mu+1}^\nu x_i}\delta_{x,y}\text{ for }\mu<\nu\\
C_{\mu\nu}&=\frac{1}{2}(C_\mu C_\nu+C_\nu C_\mu)
\end{split}
\end{equation}
The spin-flavor structure of $M_2^{\mu\nu}$ can be made explicit by
noting that
\begin{equation}
\eta_{\mu\nu}=\epsilon_{\mu\nu}\eta_\mu\eta_\nu
\qquad
\left(\epsilon_{\mu\nu}\right)_{xy}=
-\left(\epsilon_{\nu\mu}\right)_{xy}=
(-1)^{x_\mu x_\nu}\delta_{x,y}
\text{ for }\mu<\nu
\end{equation}
We have 
$\epsilon_{\mu\nu}\sim(\gamma_\mu\gamma_\nu\otimes\xi_\nu\xi_\mu)$
and, up to discretization terms, $\eta_\mu\eta_\nu C_{\mu\nu}\sim
(\gamma_\mu\gamma_\nu\otimes\mathbf{1})$. Therefore we see that indeed
$M_2^{\mu\nu}\sim(\mathbf{1}\otimes\sigma_{\mu\nu})+\mathcal{O}(a)$.

It is also straightforward to check that $M_2^{\mu\nu}$ is hermitian
and commutes with $\epsilon$. We therefore conclude that
$M_2^{\mu\nu}$ is a valid candidate for a flavor dependent mass term
and may be used for the construction of a Wilson-type operator.
\footnote{Note, that in two dimensions $M_2^{12}$ is the unique term
  required for fully lifting the twofold staggered flavor degeneracy
  (up to a trivial sign). Furthermore, one can obtain the flavor
  chiral mass term $M_1$ as the antisymmetrized product of the
  $M_2^{\mu\nu}$ via
\begin{equation}
M_1=-\frac{1}{4!}\epsilon_{\alpha\beta\mu\nu}M_2^{\alpha\beta}M_2^{\mu\nu}
\end{equation}
}

Let us first look at the operator
\begin{equation}
D_\text{k}(m_0)=D_\text{st}+r\left(2+M_1+M_2^{\mu\nu}\right)+m_0
\label{myop}
\end{equation}
which is expected to fully lift the staggered flavor degeneracy. The
choice of $\mu$ and $\nu\ne\mu$ is arbitrary and reflects the
ambiguity of defining a single flavor in the spin-flavor
representation. Note that $[M_2^{\mu\nu},\epsilon]=0$ together with the
hermiticity of $M_2^{\mu\nu}$ implies $D_\text{k}(m)\epsilon=\epsilon
D^\dag_\text{k}(m)$. Consequently, non-real eigenvalues of
$D_\text{k}(m)$ also appear in complex conjugate pairs.

In fig.~\ref{figopsmy}, the spectrum of $D_\text{k}(0)$ is plotted for
the free case. Note that the multiplicity of the real eigenmodes is
$(4,8,4)$ as compared to $(8,8)$ for the free $D_\text{A}(0)$
displayed in fig.~\ref{figopsad}, indicating that the flavor
degeneracy has indeed been lifted.

In order to investigate the symmetries of the single flavor operator
(\ref{myop}), we proceed to investigate the transformation properties
of $M_2^{\mu\nu}$ under the the staggered symmetries
\cite{Golterman:1984cy}. Note, that in addition to breaking the flavor
chiral symmetry $U(1)_\epsilon$, $M_2^{\mu\nu}$ also breaks the
discrete shift, axis reversal and hypercubic rotational
symmetries. Explicitly, we have
\begin{equation}
M_2^{\mu\nu}\rightarrow M_2^{\mu\nu}\cdot\left\{
\begin{array}{rcl}
-1& \text{for}&\rho=\mu,\nu\\
1& \text{else}&
\end{array}
\right.
\end{equation}
under both shift and axis reversal symmetries along the direction
$\rho$ and
\begin{equation}
M_2^{\mu\nu}\rightarrow R^{(\rho\sigma)}_{\mu\alpha}R^{(\rho\sigma)}_{\nu\beta}M_2^{\alpha\beta}
\label{rottran}
\end{equation}
under a rotation of the $\rho$ into the $\sigma$ direction, where the
rotation matrix $R^{(\rho\sigma)}$ is explicitly given by 
\begin{equation}
R^{(\rho\sigma)}_{\mu\alpha}x_\alpha=\left\{
\begin{array}{rcl}
x_\sigma& \text{for}&\mu=\rho\\
-x_\rho& \text{for}&\mu=\sigma\\
x_\mu& \text{else}&
\end{array}
\right.
\end{equation}
Note that rotational transformations (\ref{rottran}) can generate an
$M_2^{\alpha\beta}$ term with arbitrary $\alpha$ and $\beta\neq\alpha$
out of any given $M_2^{\mu\nu}$. This suggests that instead of adding
$M_1$ and a single $M_2^{\mu\nu}$ to the staggered operator as in
(\ref{myop}), one might instead take a more symmetric linear
combination of the different $M_2^{\mu\nu}$.

We proceed to investigate the following linear combinations
\begin{equation}
\begin{split}
M_s=\frac{1}{\sqrt{3}}\big(
&s_{12}(s_1s_2M_2^{12}+s_3s_4M_2^{34}) \\
+&s_{13}(s_1s_3M_2^{13}+s_4s_2M_2^{42}) \\
+&s_{14}(s_1s_4M_2^{14}+s_2s_3M_2^{23})
\big)
\label{coolmassterm}
\end{split}
\end{equation}
where the $s_\mu=\pm 1$ and $s_{\mu\nu}=\pm 1$ are arbitrary sign
prefactors. One can check that the effect of both shift translation
and axis inversion in $\rho$ direction is a single sign flip
$s_\rho\rightarrow -s_\rho$. Similarly, the effect of a single
hypercubic rotation is to flip the sign of one single $s_{\mu\nu}$
(cf. table~\ref{tabsmn}). As a result, $M_s$ is invariant under the
following discrete symmetries:
\begin{enumerate}
  \item
    Diagonal shift:
\begin{equation}
x\rightarrow x+\hat{1}\pm\hat{2}\pm\hat{3}\pm\hat{4}
\label{shift}
\end{equation}
  \item
    Shifted axis reversal:
\begin{equation}
x_\mu\rightarrow -x_\mu+\hat{\mu}
\label{flip}
\end{equation}
  \item
    Double rotation:
\begin{equation}
x\rightarrow R^{(\mu\nu)}R^{(\rho\sigma)}x\quad
\label{rot}
\end{equation}
    with $(\mu,\nu,\rho,\sigma)$ any permutation of $(1,2,3,4)$
\end{enumerate}
These symmetries form a subrgoup of the original discrete staggered
symmetries shift, axis reversal and hypercubic rotation
\cite{Golterman:1984cy}.

\begin{table}
\begin{tabular}{cccc|c}
\multicolumn{4}{c|}{$(\rho,\sigma)$}&sign flip\\
\hline
(1,4)&(2,3)&(3,1)&(2,4)&$s_{12}\rightarrow -s_{12}$\\
(1,2)&(3,4)&(4,1)&(3,2)&$s_{13}\rightarrow -s_{13}$\\
(1,3)&(4,2)&(2,1)&(4,3)&$s_{14}\rightarrow -s_{14}$
\end{tabular}
\caption{\label{tabsmn}Explicit table of sign flips in (\ref {coolmassterm})
under the various hypercubic rotations $R^{(\rho\sigma)}$}
\end{table}

Up to discretization effects, the flavor structure $\xi^{(s)}$ of $M_s$ is given
by a linear combination of the $\sigma_{\mu\nu}$, such that in the
common diagonal basis one can write
\begin{equation}
\xi_5=\text{diag}(1,1,-1,-1)
\qquad
\xi^{(s)}=
\left\{
\begin{array}{lll}
\text{diag}(2,-2,0,0) && \text{or}\\
\text{diag}(0,0,2,-2) &&
\end{array}
\right.
\label{fssym}
\end{equation}
which indicates, that a single $M_s$ alone should fully lift the
flavor degeneracy.
\footnote{Note that here the two flavors of one chirality
  receive an opposite mass term while both flavors of the other
  chirality remain unaffected up to discretization effects.} I
therefore propose the following symmetrized Wilson-like operator
\begin{equation}
D_\text{s}(m_0)=D_\text{st}+r\left(2+M_s\right)+m_0
\label{myops}
\end{equation}

Again, one can demonstrate that $D_\text{s}(m)\epsilon=\epsilon
D^\dag_\text{s}(m)$ so that the eigenvalues of $D_\text{s}(m)$ also
appear in complex conjugate pairs. The spectrum of the free
$D_\text{s}(0)$ is plotted in fig.~\ref{figopsmys}. One can see that
it is similar to the one of (\ref{myop}) (cf.~fig.~\ref{figopsmy})
with the same multiplicity of real eigenmodes $(4,8,4)$. Note however
that the spectrum does display a higher degeneracy of the eigenmodes
which reflects the improved symmetry properties of the free
(\ref{myops}) as compared to the free (\ref{myop}).

In the interacting theory (\ref{myops}) will receive radiative
corrections. The structure of these corrections will be restricted by
the symmetries (\ref{shift}-\ref{rot}) (in addition to the usual gauge
and baryon number symmetries). These symmetries form a subgroup of the
discrete staggered symmetries \cite{Golterman:1984cy}. The counterterm
structure of the single flavor operator (\ref{myops}) will be the one
of the 2-flavor operator (\ref{adop}) plus additional local terms that
break the hypercubic rotational symmetry but do preserve the double
rotation symmetry (\ref{rot}).
 
The dimension 3 counterterms that can appear are therefore the scalar
$\bar{\chi}\chi$, the 2-flavor mass term $\bar{\chi}M_\text{1}\chi$
and all of the $\bar{\chi}M_\text{s}\chi$. As in the case of Wilson
fermions, $\bar{\chi}\chi$ necessitates an additive mass
renormalization. To leading order, the effect of the 2-flavor mass
term $\bar{\chi}M_\text{1}\chi$ will be a relative shift of the
physical and doubler branches of the spectrum as one can see from
(\ref{fssym}). Mixing among the different $\bar{\chi}M_\text{s}\chi$
on the other hand will lead to a renormalization of the flavor
structure. This is evident from the fact that the different
$\xi^{(s)}$ do not all commute among each other. The flavor assignment
to leading order will no more be given by the $\xi^{(s)}$ but instead
by a linear combination of them.\footnote{Note however, that all
  linear combinations of the $\xi^{(s)}$ do still commute with
  $\xi_5$.} The radiative corrections will therefore modify the
particular linear combination of the four degenerate flavors of the
underlying staggered operator that will end up in the physical branch
of the spectrum. But since the details of the flavor assignment are
arbitrary in any case and do not carry any further physical
significance, the renormalized flavor assignment is as good as the
bare one and there is no need to undo this mixing. Apart from this
renormalized flavor assignment, the radiative corrections will
obviously move the relative positions of the physical and doubler
branches. This however can be absorbed by a further additive mass
renormalization as long as a clear separation between the physical and
doubler branches is maintained (which is expected to be the case
outside the strong coupling regime).

Note that radiative corrections are generally expected to be
suppressed by UV-filtering (and Symanzik improving) the fermion
operator in much the same way as it is the case for staggered
\cite{Blum:1996uf,Orginos:1999cr} and Wilson fermions
\cite{DeGrand:1998mn,Capitani:2006ni}.

A further detailed investigation of this issue is beyond the scope of
this paper but will be an essential ingredient in judging the practical
usefulness of (\ref{myops}).

We now proceed to use $D_\text{s}(m)$ as the kernel of an overlap
operator \cite{Neuberger:1997fp}. We define a massless overlap
operator
\begin{equation}
D_1=\rho\left(1+\epsilon \text{sign}(\epsilon D_\text{s}(-\rho))\right)
\label{myovop}
\end{equation}
that is conjectured to describe a single flavor if $\rho$ is chosen
properly, i.e. in a range that extends to $\rho\in(0,2)$ in the free
case. Whether (\ref{myovop}) constitutes a valid overlap operator
remains to be seen. Specifically, the counterterm structure and the
question of locality and a spectral gap \cite{Golterman:2003qe} in the
hermitian kernel operator $\epsilon D_\text{s}(-\rho)$ require further
investigations.

Beyond the remaining conceptual questions it is interesting to
speculate about the usefulness of either (\ref{myops}) or
(\ref{myovop}) for lattice QCD calculations. The obvious advantage of
(\ref{myops}) as compared to a standard Wilson operator are its
smaller size (by a factor of 4) and the reduced condition number of
$D^\dag D$ (another factor of 4 for $m_0=0$ in the free case). This
leads to a reduction in storage requirement by a factor of 4 and to a
naively estimated speedup factor of $\sim 16$ for conjugate gradient
inversions or the construction of the overlap operator (\ref{myovop}).
Note that in contrast to the staggered fermion case these improvements
originate in the huge reduction of UV modes (by eliminating 12 of the
15 doubler fermion species) and do not come at the price of
quadrupling the IR modes in the physically relevant branch of the
spectrum.

There are also two obvious disadvantages of (\ref{myops}) as compared
to a standard Wilson operator. The first one is the appearance of
two-hop terms in the Wilson-like operator (\ref{myops}). These terms
could lead to a slower numerical implementation when compared to
strictly one-hop operators. It is however interesting to note, that
the additional gauge links needed for the two-hop terms are exactly
the same gauge links one needs for the construction of a clover-term
\cite{Sheikholeslami:1985ij} that can be used to $O(a)$ improve a
Wilson operator. One therefore expects on the one hand that the speed
loss due to 2-hop terms would be approximately equal to the speed loss
by including a clover term (which is typically about $30\%$) and on
the other hand that the inclusion of a clover term into (\ref{myops})
would essentially be free.

A second concern is that (\ref{myops}) was constructed using the
staggered spin-flavor basis and will therefore inherit the nonlocal
definition of the spin matrices. In contrast to staggered fermions
however, there will be no significant flavor mixing due to the
Wilson-like lifting of the flavor degeneracy. Doubler states will
quickly die out in correlation functions and will not have to be
disentangled as for staggered fermions. I therefore do not expected
that e.g. ground state hadron masses or simple matrix elements like
decay constants or $B_K$ will be harder to obtain than with Wilson
fermions. It might however turn out to be difficult to extract some
short distance observables with nontrivial spin structure in this
formulation.

\section*{Acknowledgments}
I would like to thank Maarten Golterman and David Adams for valuable
comments on the manuscript and Stephan D\"urr for helpful
discussions. In reaction to a previous version of this manuscript I
was informed that Philippe de Forcrand, Aleksi Kurkela and Marco
Panero have presented a spectrum plot of a single flavor staggered
operator at a recent workshop \cite{DeForcrand}. Although no further
details were given, the plot closely resembles
fig.~\ref{figopsmy}. This work was supported by the DFG grant SFB-TR
55.

\bibliographystyle{elsarticle-num}

\end{document}